\documentclass[aps,prl,onecolumn, 11pt,showpacs,superscriptaddress]{revtex4-1}
\usepackage{hyperref}
\usepackage{graphicx}
\usepackage{amsmath}
\usepackage{amssymb}
\usepackage{color}

\usepackage{dcolumn}
\usepackage{bm}
\usepackage{color}

\usepackage{tabularx}
\usepackage{multirow}
\usepackage{selinput}
\usepackage{array}
\newcolumntype{L}[1]{>{\raggedright\let\newline\\\arraybackslash\hspace{0pt}}m{#1}}
\newcolumntype{C}[1]{>{\centering\let\newline\\\arraybackslash\hspace{0pt}}m{#1}}
\newcolumntype{R}[1]{>{\raggedleft\let\newline\\\arraybackslash\hspace{0pt}}m{#1}}

\begin{document}

\title{Network diffusion capacity unveiled by dynamical paths.}

\author{Tiago A. Schieber}
\affiliation{%
Departamento de Ci\^encias Administrativas, \\Universidade Federal de Minas Gerais, Belo Horizonte, MG, Brazil}%

\author{Laura C. Carpi}
\affiliation {Instituto Nacional de Ci\^encia e Tecnologia, Sistemas Complexos, INCT-SC, CEFET-MG, Belo Horizonte, Brazil}%
\affiliation{Machine Intelligence and Data Science Laboratory (MINDS), \\Universidade Federal de Minas Gerais, UFMG
31270-000 Belo Horizonte, Brazil}%
\author{Panos M. Pardalos}
\affiliation{Industrial and Systems Engineering, University of Florida, Gainesville, FL, USA }%

\author{Cristina Masoller}
\affiliation{Departament de F\'isica, Universitat Polit\`ecnica de Catalunya. Rambla St. Nebridi 22, Terrassa 08222, Barcelona, Spain}%

\author{Albert D\'iaz-Guilera}
\affiliation{
Departament de F\'isica de la Mat\`eria Condensada, Universitat de Barcelona, Barcelona, Spain
}%
\affiliation{Universitat de Barcelona. Institute of Complex Systems (UBICS), 08028 Barcelona, Spain}

\author{Mart\'in G. Ravetti}
\email{martin@dcc.ufmg.br}
 \affiliation{%
Departamento de Ci\^encia da Computac\~ao, \\Universidade Federal de Minas Gerais, Belo Horizonte, MG, Brazil
}%
\date{\today}

\begin{abstract}

Improving the understanding of diffusive processes in networks with complex topologies is one of the main challenges of today's complexity science. Each network possesses an intrinsic diffusive potential that depends on its structural connectivity. However, the diffusion of a process depends not only on this topological potential but also on the dynamical process itself. Quantifying this potential will allow the design of more efficient systems in which it is necessary either to weaken or to enhance diffusion. Here we introduce a measure, the {\em diffusion capacity}, that quantifies, through the concept of dynamical paths, the potential of an element of the system, and also, of the system itself, to propagate information.  Among other examples, we study a heat diffusion model and SIR model to demonstrate the value of the proposed measure. We found, in the last case, that diffusion capacity can be used as a predictor of the evolution of the spreading process. In general, we show that the diffusion capacity provides an efficient tool to evaluate the performance of systems, and also, to identify and quantify structural modifications that could improve diffusion mechanisms.

\end{abstract}

\maketitle

The diffusion of information, with varying levels of complexity, is ubiquitous in our everyday life. Advancing our understanding of diffusive processes is a fundamental challenge with critical practical applications across a wide range of spatial scales. Diffusion magnetic resonance is an imaging technique that allows studying the brain structural and functional connectivity~\cite{Hagmann2007,Shouliang2016}. A diffusion-like process describes the action of infectious agents that attack our immune system spreading as fast as they can~\cite{Chen2017}. 
On the large scale, the billions of individuals commuting between different geographical regions daily, constitute the highly complex global human mobility system~\cite{Morita2016,Scarpino2019,Kraemer2019}.
Similarly, gossip spreads through vast complex social networks~\cite{Kempe2003,Lappas2010,Qi2018,Shao2018,Bovet2019,Pierri2020,Zhou2020}. These diffusive phenomena, present in our everyday life, have motivated research to understand the mechanisms that enhance or suppress diffusion and quantify their impacts~\cite{Myers2012,Akbarpour2018,Arruda2020,Darbon2021,Iacopini2020}.

The analysis of diffusive processes usually assumes that interaction networks represent their average behavior~\cite{Myers2012}; some models consider random navigation with several levels of information~\cite{Beekman1967,Havlin1987,bouchaud1990,Domenico2017,Giuggioli2020,Bertagnolli2021}, and others, consider the topological shortest paths~\cite{Fornito2016}.
However, as diffusion occurs over all the existing paths in the network, the more detailed the information extracted from the network connectivity, the more precise the understanding about the behavior of dynamical processes flowing through it~\cite{Hens2019,Gonzales2019,Bertagnolli2021,Boguna2021}.


To advance the understanding of the diffusion of dynamical processes in complex networks, we propose a process-dependent measure, the diffusion capacity, that takes into account information of the network structural connectivity~\cite{Schieber2017,Carpi2019} and considers the dynamical process that diffuses in the network. It is essential to highlight that we refer to the network structural connectivity or topology, in the sense of its structural binary information disregarding the dynamics occurring on the system, for which two nodes can be connected or not to each other. Weights associated with the links contain information of the dynamical process that can be either parameter of an explicit dynamical equation (e.g., coefficient of heat transfer), a result of a sampling process (e.g., the average time in a transportation network), a probability value (e.g., probability of infection), among others.

\begin{figure}[h]
 \centering
 \includegraphics[width=\linewidth]{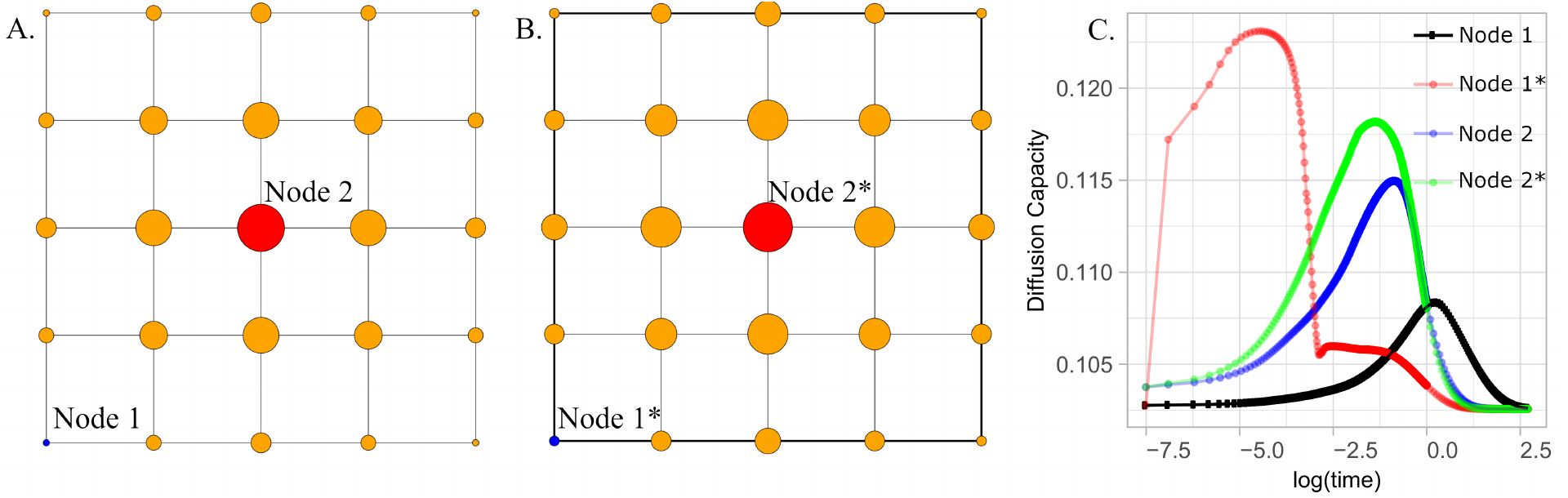}
 \caption{Regular grid of 25 nodes in which their sizes correspond to their diffusion capacity values when no flow is present (A), and the same grid in which peripheral links posses a higher weight (B). Two different heat diffusion process are initiated in nodes 1 and 2 in both structures, by the assignation of a temperature that is $25$ degrees higher than the temperature of the orange nodes. The time evolution of the diffusion capacity values of the system are shown in (C). }
 \label{fig:grid}
\end{figure}

We begin with a simple and intuitive example to illustrate the concept of the diffusion capacity.
We consider a heat diffusion model~\cite{Thanou2017}  where a system is composed of a regular grid, and where each node $i$, has a given initial temperature, $x^0_i$ (the model equations and parameters are presented in section A of the SI). Nodes will interact by exchanging heat with a speed proportional to their temperature difference and the thermal conductivity, to finally reach thermal equilibrium. In thermal equilibrium, the diffusion capacity of nodes (represented by the size in Figure \ref{fig:grid}-A) depends exclusively on the topology, as there is no energy flow. Then, more central nodes have higher capacity as they are connected to a higher number of nodes. However, the diffusion capacity of the nodes varies in time when the system is out of equilibrium and energy flows. As an example, we consider an initial condition in which all nodes have the same temperature except one node (either 1 or 2 in Figure \ref{fig:grid}-A), which has a higher temperature. In this situation, a diffusion process starts, and the diffusion capacity of the system varies in time as shown in Figure \ref{fig:grid}-C. We see that the diffusion capacity of the system changes as the process evolves to reach thermal equilibrium, increasing until a maximum value that occurs when the temperature heterogeneity is highest, and then decreasing to reach thermal equilibrium. The maximum value is higher when the process initiates in node two, because this node (being located in the center of the network) possesses a higher topological diffusion capacity than the peripheral, allowing the system to reach thermal equilibrium faster.
Now, if the thermal conductivity of peripheral links (considered as a weight) is increased, as shown in Figure \ref{fig:grid}-B, the diffusion capacity of all nodes increases, as well as the speed of the heat transfer. Then, the highest diffusion capacity is reached when node 1 is perturbed as shown in Figure \ref{fig:grid}-C. This configuration of Figure \ref{fig:grid}-B is also faster Figure \ref{fig:grid}-A, to reach thermal equilibrium (red line).

\subsection{Diffusion capacity}

Let $G=(V,E,W)$ be a weighted network composed by a set of vertices, V, and edges, E, whose weights W, contain information about the specific dynamic process, D.
Weights allow us to associate to each link a distance that is the inverse of the link's weight.
In this situation, the most efficient structure for diffusing process D is a fully connected structure whose links have infinite weights. Therefore, we consider a fully connected graph with infinite weights as a ``reference graph", $G_{ref}$, and define the diffusion capacity of $G$, $\Lambda_D(G)$, as the distance between $G$ and $G_{ref}$. For that, we use a measure of dissimilarity between graphs $G$ and $G'$~\cite{Schieber2017}, based on the distance between two sets of probability distributions associated to them: the dynamic node distance distributions (dNDDs).
The definition of dNDD of node $i$ in $G$, $\mathbb P_i(G)$, is presented in Methods and it takes into account both, the topological shortest paths and the weighted shortest paths to node $i$. Then, the diffusion capacity of node $i$, $\Lambda_i(G)$, is the inverse of the distance between the dNDD of node $i$ in $G$ and the dNDD of node $i$ in $G_{ref}$. We measure the distance between these distributions using the cumulative Jensen-Shannon divergence (CDD, described in section E ). Therefore, the diffusion capacity of node $i \in G$, $\Lambda_i(G)$, is


 \begin{equation}
 \Lambda_i(G)=[CDD(\mathbb P_i,\mathbb P_{ref})]^{-1}
 \end{equation}

and the diffusion capacity of the whole network $G$, $\Lambda_D(G)$, is the average over all nodes
 \begin{equation}
  \Lambda(G)=\frac{1}{|V|}\sum_{i\in V}\Lambda_i(G).
 \end{equation}

\begin{figure}[htb]
 \centering
 \includegraphics[scale=0.8]{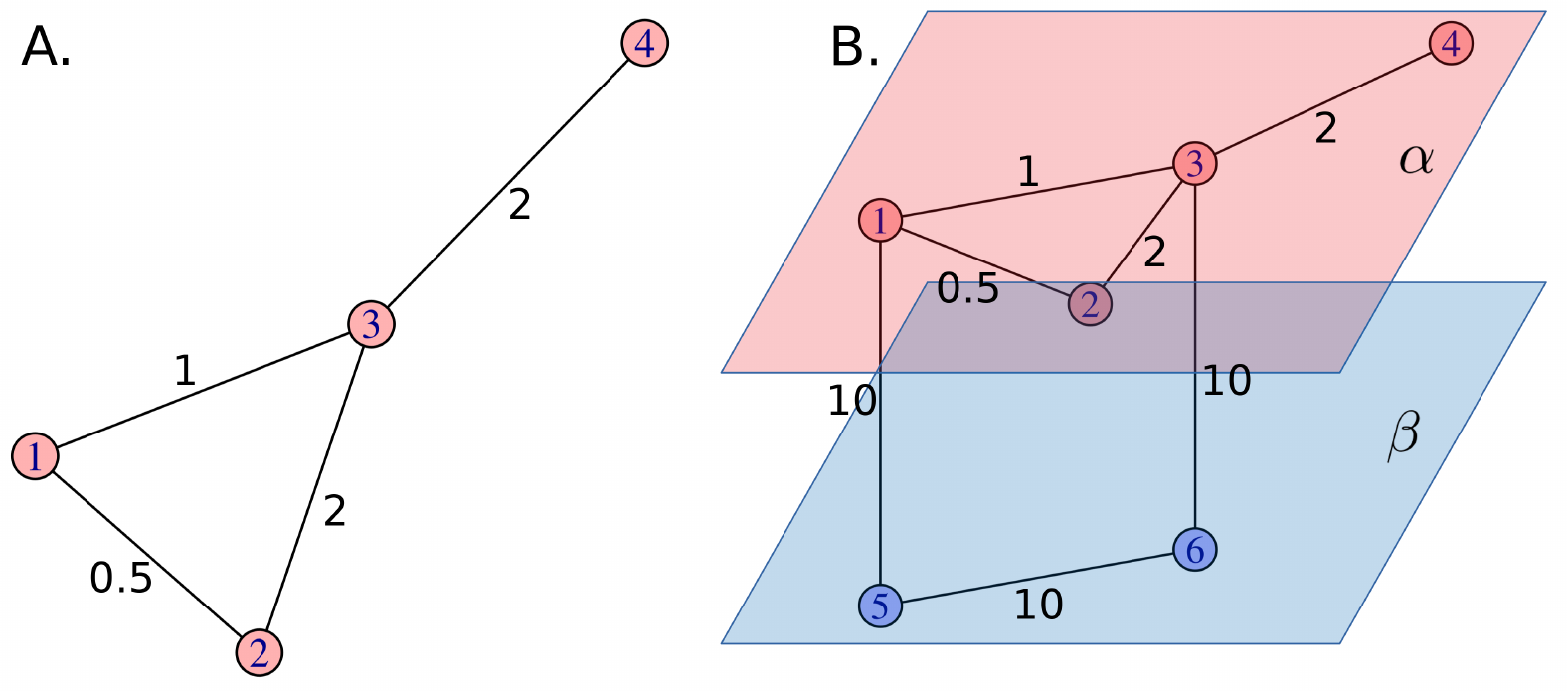}
 \caption{\label{fig:small} Single weighted network composed by 4 nodes (A), and a multilayer network composed by two layers, in which the layer $\alpha$ is the network in A, but connected to layer $\beta$ composed by two other nodes (B). In (A) we note that the topological shortest path between nodes $1$ and $2$ is directly by link $1\rightarrow 2$ ($D_g=1$). The weight of this link is $0.5$, then the corresponding weighted distance is $2$ ($D_w=2$). However, if instead of using this direct path $1\rightarrow 2$, is used the two-steps path $1\rightarrow 3\rightarrow 2$ ($D_g=2$), the weights are $1$ ($D_w=1$) for link $1\rightarrow 3$ and $2$ ($D_w=1/2$) for link $3\rightarrow 2$. Then the weighted distance of the complete path correspond to the sum $D_w=1+1/2=3/2$. In this particular case, the topological shortest path does not coincide with the weighted shortest path. However, we can see in (B) that going from $1$ to $2$ through layer $\beta$, the weighted shortest path corresponds to $D_w=1/10+1/10+1/10+1/2=8/10$, being more advantageous.}
 \end{figure}
Figure \ref{fig:small}-A displays a simple example to illustrate the application of Eq. (1): node 3 ($\Lambda_3(G)=1.48$) is the one with the highest diffusion capacity, which reflects its central location and the weights of its links.  However, if a process needs to start in one of the peripheral nodes, which one would be the more convenient? Our analysis indicates that node 2 ($ \Lambda_2(G)=0.48$) has larger diffusion capacity than nodes 1 ($ \Lambda_1(G)=0.43$) and 4 ($ \Lambda_4(G)=0.38$). This decreasing order of the node's diffusion capacity values corresponds to the decreasing similarity of each node's distance distribution with the reference. The situation presented in Figure~\ref{fig:small}-B in which paths between nodes in $\alpha$ goes through layer $\beta$, is described later.

\begin{figure}[h]
 \centering
 \includegraphics[width=\linewidth]{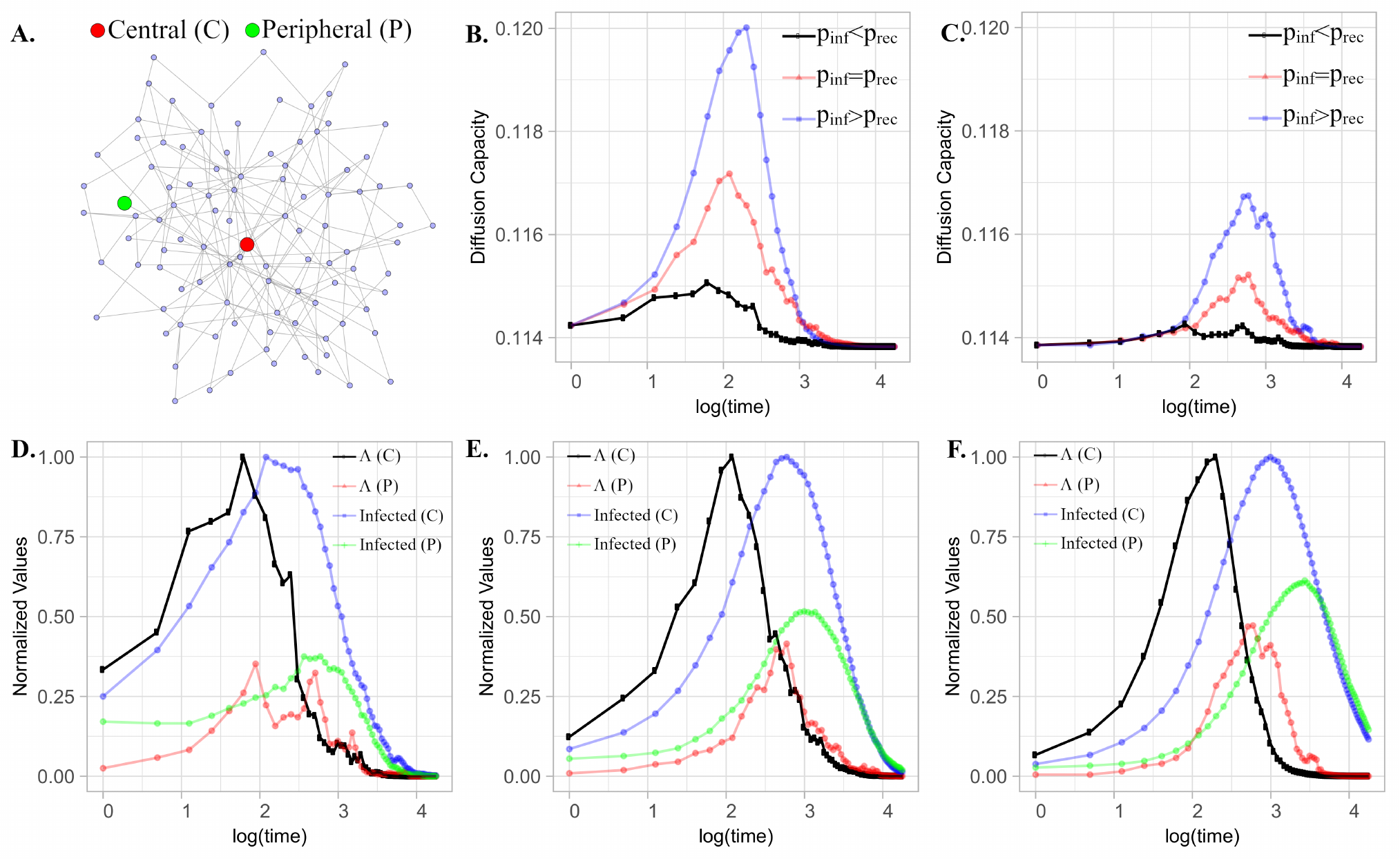}
 \caption{Small network highlighting a central (red) and a peripheral (green) node (A). Network diffusion capacity evolution of a SIR model when the central node (B) and a peripheral node (C) is initially infected, for three different infection probabilities. The three remaining figures show mean values of diffusion capacities and infected individuals of $100$ realization of SIR processes initiated in central and peripheral nodes for $p_{inf} < p_{rec}$ in (D), $p_{inf} = p_{rec}$ in (E) and $p_{inf} > p_{rec}$ in (F). Here, we consider a constant $p_{inf}=0.1$ and $p_{rec}=0.05,\;0.10\;\mbox{and}\;0.20$.}
 \label{fig:sir3}
\end{figure}

Considering the SIR model~\cite{kermack1927,Bartoszynski1965,Bacaer2011}, a simple mathematical model for the spread of epidemic diseases, in which, nodes can be susceptible, infected, or recovered from an infectious agent.
Recovered nodes are immune to the disease, while a susceptible node can become infected if it is in contact with an infected node.
We show here, the evolution of the SIR model in a small network (Figure \ref{fig:sir3}-A) when one central node is initially infected, and also when a peripheral node initiates the epidemic process. In this experiment, we consider the probability of a susceptible node to become infected is $p_{inf}=0.1$ and different recovery rates $p_{rec}$ values.

In general, when the process begins with the infection of a central node, the number of new infections grows more and faster, reaching earlier immunity, than when the process is initiated in a peripheral node. In this case, the number of new infected nodes grows less and slower, having its maximum later. The evolution of the dynamic also depends on the probability of recovery ($p_{rec}$) of infected nodes, that induces different outcomes.

In Figure \ref{fig:sir3} it is depicted a SIR model when the probability of infection is greater than the probability of recovery ($p_{inf} > p_{rec}$), when the probabilities are equal ($p_{inf} = p_{rec}$), and finally when the probability of infection is lower than the probability of recovery ($p_{inf} < p_{rec}$). The lower the probability of recovery, the faster the disease spreads. Figures \ref{fig:sir3}-B and C show the evolution of the diffusion capacity when the central, and peripheral nodes are initially infected for different recovery probability values. For $p_{inf} > p_{rec}$  values of diffusion capacity present the highest peak, lower for the process initiated in the peripheral node. The lowest values correspond to the case of $p_{rec}>p_{inf}$. It is interesting noticing that when process is initiated in the peripheral node, the diffusion capacity values show two small and similar peaks corresponding to the initial increase of infected nodes, and to the infection of the central nodes which have a high topological diffusion capacity.

Figures \ref{fig:sir3}-D, E, and F compare the different infection strategies explained above. These figures show that the initial diffusion capacity is higher when the process is initiated in the central node. As the process evolves, the diffusion capacity increases until a maximum that appears earlier than the maximum number of cases, showing itself as an early indicator of the peak of the epidemic process. This information can be useful to plan strategies to reduce the impact of these kind of spreading diseases.

We then generalize the measure for interconnected structures (multilayer networks), which considerably increases the system's complexity. Figure \ref{fig:small}(B) depicts a small example of a multilayer system composed by two networks. It can be seen that, in this case, going from node $1$ to node $2$ in four steps through layer $\beta$, is advantageous than staying in layer $\alpha$ using path $1-3-2$ or directly $1-2 $, as their weighted shortest path are respectively $D_w=8/10$, $D_w=3/2$, and $D_w=1$.
In a multilayer system, if there are no interlayer connections, diffusion occurs independently in each layer, and the diffusion time is determined by the slowest one. However, when the interlayer strength is low, the diffusion time may become excessively long, as these weak interlayer connections slow down the dynamics of both layers. On the other hand, a strong interaction between layers enhances diffusion.

For the case of diffusion processes in multiplex networks, which are multilayer networks with the restriction of having the same set of nodes in all layers, and intralayer connections exclusively through the same nodes, it has been found that the time for the network to reach equilibrium, scales with the inverse of the smallest positive eigenvalue of the Laplacian matrix. It was also found that some diffusion processes present not trivial behaviors such as super-diffusion, by which the multiplex structure reaches a steady-state faster than any of the layers in isolation~\cite{Gomez2013}. This approach has been successfully used in several applications~\cite{Arenas2008,Reza2007,Nishikawa2010,Constantino2019,Cencetti2019} and later expanded to other phenomena~\cite{delGenio2016} such as synchronization and reaction-diffusion processes~\cite{Asllani2014,Asllani2014b,Kouvaris2015,Busiello2018}. Diffusion times computed by diffusion capacity are in excellent agreement with those determined by the Laplacian matrix's smallest positive eigenvalue (see section B of the SI).

 Let $\vec G=(G_1,G_2,\dots,G_M,\mathbb{E})$ a multilayer weighted network, we define, for each node $i\in V_L$ and for all $L=1,2,\dots,M$:
 \begin{enumerate}
  \item The Node Diffusion Capacity (${\cal M}_i$):
 \begin{equation}\label{mlie}
  {\cal M}_i(\vec G)=\left[\frac{1}{2}CDD(\mathbb P_{i},\mathbb P_{ref})+\frac{1}{2M-2}\sum^M_{\beta=1,\beta\not= L}CDD(\mathbb P^\beta_i,\mathbb P_{ref})\right]^{-1}
 \end{equation}

 \item The Layer Diffusion Capacity is defined as the average of ${\cal M}_i$ over all the nodes in a layer,
  \begin{equation}
 \label{TML}
  {\cal M}(G_L)=\frac{1}{|V_L|}\sum_{i\in V_L}{\cal M}_i(G)
 \end{equation}

\item The Multilayer Diffusion Capacity is defined as the average of  ${\cal M}(G_L)$ over all the layers, weighted by the layer sizes:
  \begin{equation}
 \label{MDC}
  {\cal M}(\vec G)=\frac{\displaystyle \sum_{L=1}^M|V_L|{\cal M}(G_L)}{\displaystyle\sum_{L=1}^M|V_L|}
 \end{equation}

 \end{enumerate}

In the first term of Eq.~(\ref{mlie}), $\mathbb{P}_{i}$ is the distribution of the multilayer dynamical paths between node $i$ and the other nodes of layer $G_L$ (see methods). $CDD(\mathbb P_{i},\mathbb P_{ref})$ quantifies the diffusive potential this node has, as a consequence of its connectivity configuration in the multilayer structure as a whole, through a distance to a reference distribution. In the second term,  $\mathbb P^\beta_i$ represents the distribution of the multilayer dynamical paths between node $i$ and the other nodes of layer $G_L$, for which paths are imposed to go through layer $\beta$ (see methods). The average of $CDD(\mathbb P^\beta_i,\mathbb P_{ref})$ captures the effect caused by the presence of all $\beta \neq L$, on node $i$. In this way, the multilayer diffusion capacity of node $i$, ${\cal M}_i(\vec G)$ is represented by the structural and dynamical dissimilarity between the multilayer connectivity of node $i$ in $G$, and the multilayer connectivity of node $i$ in $G_{ref}$.

\begin{figure}[h]
 \centering
 \includegraphics[width=\linewidth]{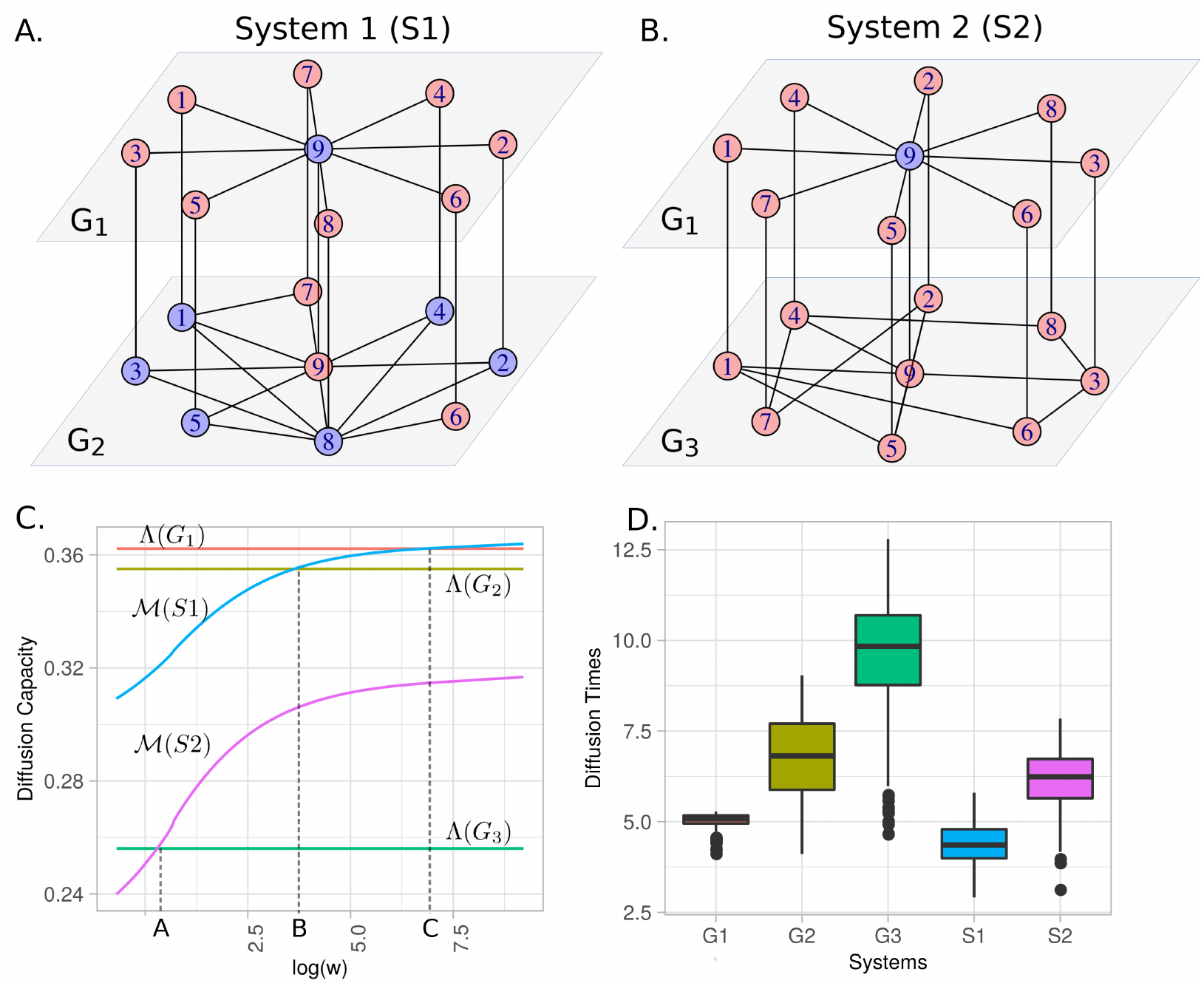}
 \caption{(A) and (B) depict two duplex networks ($S_1$ and $S_2$). (C) shows the Diffusion Capacity $\Lambda$ of  $G_1$, $G_2$ for different weights (logarithmic scale), and $G_3$ and Multilayer Diffusion Capacity $\cal M$ of $S_1$ and $S_2$. (D) Average diffusion times of the heat model described in section SI-D considering 1000 seeds  ($95\%$ C.I.). Nodes in blue/red represent those that decrease/increase their diffusion capacity in the multilayer structure, compared to their diffusion capacity in isolation.}
 \label{fig:smallest}
\end{figure}

Figure \ref{fig:smallest} shows two duplex systems containing nine nodes in each layer, coupled with a constant interlayer weight. System $S_1$ is composed by layers $G_1$ and $G_2$ (Figure~\ref{fig:smallest}-A), and system $S_2$ is composed by layers $G_1$ and $G_3$ (Figure~\ref{fig:smallest}- B). Figure \ref{fig:smallest}-C depicts the diffusion capacities values for isolated layers (${{\Lambda}(G_1)}$, ${\Lambda}(G_2)$ and ${\Lambda}(G_3)$), and for the multilayer systems, (${\cal M}(S1)$ and ${\cal M}(S2)$), for different interlayer weights.

Due to the high centrality of node nine in layer $G_1$, ${\Lambda}(G_1)>{\Lambda}(G_2)$ and ${{\Lambda}(G_1)>{\Lambda}(G_3)}$. For a small interlayer strength value, ${\cal M}(S1)$ and ${\cal M}(S2)$ are lower than all layers diffusion capacity values. Increasing interlayer weights, the diffusion capacity of the system $S_1 $ goes, from point B, the Diffusion Capacity of layer $G_2$, and the Diffusion Capacity of the system $S_2$ goes, from point A, the diffusion capacity of layer $G_3$. Continue increasing the strength of interlayer weights, ${\cal M}(S1)$ also reaches ${{\Lambda}(G_1)}$, going it from point C, marking the transition from where ${\cal M}(S1)$ becomes higher than the diffusion capacity of every isolated layer.  At this point, it is possible to say that all layers somehow gain due to the multilayer topology as ${\cal M}(G_1)>{\Lambda}(G_1)$, ${\cal M}(G_2)>{\Lambda}(G_2)$ and ${\cal G}(S1)> 1$. Instead, in $S2$ $\Lambda(G_1)$ is not reached by ${\cal M}(S2)$.

By comparing the node diffusion capacities $\Lambda_{i}$ and ${\cal M}_{i}$, it is possible to know if it is advantageous or not for a node to be coupled with another network. By comparing the diffusion capacities $\Lambda$ and ${\cal M}(G_L)$, it possible to know if it is advantageous or not for a layer, in terms of diffusion, to be part of a multilayer system. Node nine in layer $G_1$,  for example, reduces its diffusion capacity when it is part of $S1$, and S2. The same is valid for nodes one to five and eight in layer $G_2$. The rest of the nodes improve their diffusion capacities (see table~\ref{fig:tabela} of the Sl).

To better explore this relationship between single and multilayer diffusion capacities, we define the relative gain ${\cal G}$, as the ratio between the system's diffusion capacity and the highest diffusion capacity of the isolated layers:
\begin{equation}
\label{R}
{\cal G}={\cal M}(\vec G)/ max( {\Lambda(G_1)},{\Lambda(G_2)},\dots,{\Lambda(G_M)})
\end{equation}
In this way, ${\cal G}$ quantifies the improvement of the diffusion capacity value of the system, in relation to the diffusion capacity of its most diffusive isolated layer.

\begin{figure}[h]
 \centering
 \includegraphics[width=\linewidth,keepaspectratio=true]{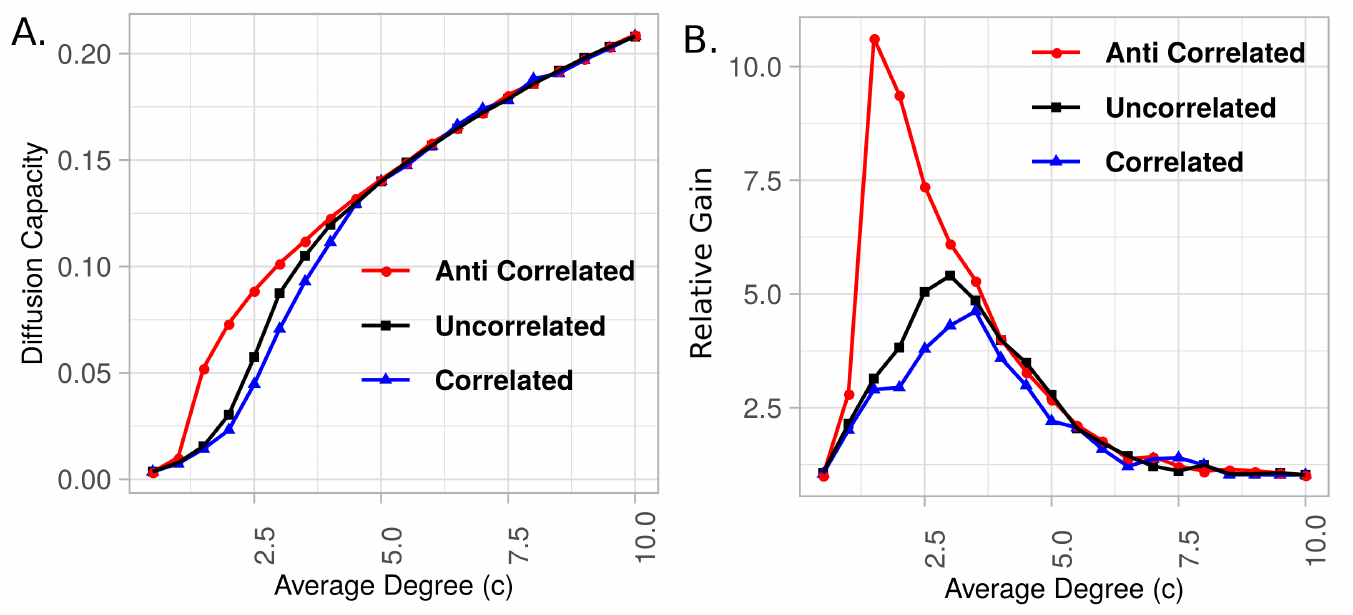}
 \caption{Average diffusion capacity values (A) and relative gain $\cal G$ values (B) of 100 realization of duplex networks considering three different inter-layers connectivity strategies. Readers should refer to SI-E section for and the corresponding heat maps of negatively correlated, uncorrelated and positively correlated layers.}
 \label{fig:control}
\end{figure}

Let us consider multilayer structures composed by two layers (duplex networks), whose structures are constructed in three different ways. The first one considers that both layers are random, uncorrelated, and undirected graphs, characterized by Poisson distributions with the same mean value. For the second structure, we select replica nodes by degree correlation, that is, nodes with more similar degrees in the different layers, are more likely to be replicas (positively correlated), and for the third structure, replicas nodes possess more dissimilar degrees in the different layers (negatively correlated)~\cite{Nicosia2015}.
Figure \ref{fig:control} shows the average diffusion capacity values of 100 realizations for the systems above mentioned, for different average degree values. The higher diffusion capacity values correspond to the system constructed by negatively correlated nodes, followed by the structure with the random selection, and finally, the system constructed by the positively selected nodes. As shown in Figure~\ref{fig:control}-A, this difference is higher for low average degrees, becoming more similar as the average degrees increase. The same behavior is valid for the average diffusion capacity values for isolated layers; however, the difference is more accentuated. Figure \ref{fig:control}-B shows the relative gain $\cal G$. In the first case, we observe that, as the average degree of the layers increases, ${\cal G}$ increases due to the lack of mutually connected components, as the shortest paths correspond, in the majority, to interlayer links. This behavior remains until mutually connected components emerge, revealing a phase transition, and determining the value of the average degree $c$ from which ${\cal G}$ stars to decrease. This result is consistent with findings in~\cite{Menichetti2016}, where a hybrid phase transition with a discontinuity in the number of driver nodes is observed for the same $c$ value. The relative gain $\cal G$ of the negatively correlated system possess the highest value, followed by uncorrelated system and then by the positively correlated one that shows two local maxima values, one at a small average degree value $c$ and a more pronounced one at an intermediate $c$ value.  All systems possess similar $\cal G$ values for intermediate to high $c$ values. Another interesting example is presented in section G of the SI.

\subsection{Conclusion}

Concluding, in this work we have introduced the concept of diffusion capacity, a quantity that reflects the potential of a network element, or a system, to diffuse information. To define this concept, and considering that diffusion processes use all existing paths, we propose a method to build a probability distribution that encode information about topological characteristics of the structure, and also dynamical features of the diffusion process itself. We call it, the dynamical node distance distribution because characteristics of the dynamical process are included as weights in the links, without excluding the topological binary information.

It can be used in a wide variety of situations, that include diffusion processes such as the diffusion of energy or epidemic processes, and information flows such as gossip or news spreading in social networks. It allows to characterize the dynamical changes of the capacity to diffuse of the network elements, or system, detecting time and conditions to reach a desired situation.

Diffusion capacity is also defined for multilayer systems, and a quantity called  relative gain is proposed to measure the gain/loss in diffusion capacity a network element has, for being part of the multilayer structure.
By analyzing duplex networks, we found that heterogeneous coupling, regarding the node's degree, enhances diffusion processes.
This analysis allows the identification of connectivity configurations that could improve the diffusion capacity of the of system, turning these measures, strong candidates for the evaluation of optimization strategies to design more efficient networks.

\subsection{Methods: Dynamical node distance distribution (dNDD)}

To define the diffusion capacity, we first construct what we call a dynamical node distance distribution (dNDD), that quantifies, the costs, gains and loses resulting for following the topological or weighted shorted path, to reach node $y$ from node $x$. The dNDD takes into account a trade-off between the topological shortest paths and the weighted shortest paths, where the weights are related to a particular dynamical process that diffuses in the network, and allows associating to each link a distance that is the inverse of the link's weight. In this way, a rich information about the network connectivity is included in the probability distribution.


\subsubsection{Dynamical paths in a single network}

The geodesic distance between two nodes, $D_g(i,j)$, is the minimum number of links separating them, while the weighted distance, $D_W(i,j)$, is the minimum sum of the links' distances. As shown in Fig.~\ref{fig:small}(a), the topological shortest paths do not always coincide with the weighted shortest paths. To quantify the effect of the presence of both shortest paths, we propose the quantity $\Delta_{i,j}=D_W(i,j)/D_g(i,j)$ as the net difference between the topological shortest path between nodes $i$ and $j$, instead of using the weighted shortest path. 

If $\Delta_{i,j}<1$ ($D_g > D_W$) the weighted shortest path connecting $i$ to $j$ is more advantageous than the topological shortest path, and $1-\Delta_{i,j}$ quantifies the gain.

If $\Delta_{i,j}>1$, $D_g(i,j) < D_W(i,j)$ then the weighted shortest path connecting $i$ to $j$ is less advantageous than the topological shortest path, and $1-1/\Delta_{i,j}$ quantifies the loss.

Let $G=(V,E,W)$ be a weighted network (directed or not), where $V$ is the set of nodes, $E$ the set of links and, $W$ the set of the weights.

For each $i\in V$ and $d=1,2\dots|V|-1,\infty$, let $\Gamma_i(d)=\{y\in V\;|\;D_g(i,j)=d\}$. We define the fraction of disconnected nodes from $i$, as $\mathbf{p}_i(\infty)=|\Gamma_i(\infty)|/(|\mathbf{V}|-1)$, and the fraction of nodes at geodesic distance $d$ from $i$ for $d\not=\infty$, $\mathbf{p}_i(d)=|\Gamma_i(d)|/(|\mathbf{V}|-1)$ as the triple:
$$
[\mathbf{p}^+_i(d),\mathbf{p}^0_i(d),\mathbf{p}^-_i(d)]= \frac{1}{|V|-1}\sum_{y\in \Gamma_i(d)}
\left[\max(1-\Delta_{i,j},0),\min\left(\Delta_{i,j},\frac{1}{\Delta_{i,j}}\right),\max\left(1-\frac{1}{\Delta_{i,j}},0\right)\right].
$$

in which, $\mathbf{p}^+_i(d)$ is the fraction of nodes at geodesic distance $d$ from $x$, for which their weighted shortest path is smaller than $d$, $\mathbf{p}^-_i(d)$ indicates the opposite situation, and $\mathbf{p}^0_i(d)$ is the fraction of nodes for which there is no gain or loss in following either paths. Then, if all weights are one, $\mathbf{p}^+_i(d)=0$ and $\mathbf{p}^-_i(d)=0$.
In the extreme case that weights tend to infinity  $\mathbf{p}^+_i(d)\rightarrow |\Gamma_i(d)|/(|V|-1)$, $\mathbf{p}^-_i(d)=0$ and $\mathbf{p}^0_i(d)\rightarrow 0$, being the topological configuration, irrelevant.
On the other hand, if weights tend to zero, $\mathbf{p}^-_i(d)\rightarrow |\Gamma_i(d)|/(|V|-1)$, $\mathbf{p}^+_i(d)=0$ and $\mathbf{p}^0_i(d)\rightarrow 0$, approximating to the situation of considering only the geodesic distances.

The {\em dNDD} is then defined, for each node, as:
\begin{equation}
 \mathbb{P}_i=[\mathbf{p}^+_i(1),\mathbf{p}^0_i(1),\mathbf{p}^-_i(1),\dots,\mathbf{p}^+_i(|V|-1),\mathbf{p}^0_i(|V|-1),\mathbf{p}^-_i(|V|-1),\mathbf{p}_i(\infty)],
 \label{eq:ndwf}
 \end{equation}
In this way, the most efficient structure, that correspond to a fully connected network with whose links have infinite weights, is represented by a {\em dynamical NDD} with the form $\mathbb P=(1,0,0,\dots,0)$. As a simple numerical example, we present in section C of the SI, the complete computation of the {\em dNDD} of the network in Figure~\ref{fig:small}(A).

\subsubsection{Dynamical paths in a multilayer network}

The evaluation of the shortest path between two nodes, present in the same layer of a multilayer structure, contemplates the paths of the corresponding layer, and also paths that go through the others. To study them, we use the concepts of single and doubly connected nodes, proposed in the lace expansion method~\cite{Hara1990}. We define that two nodes $x$ and $y$ are {\em doubly connected} ($x \Leftrightarrow y$) if it exists, between them, at least a shortest path going through, at least, two different layers. Then, two nodes are {\em single connected} ($x\leftrightarrow y$) if they are not doubly connected. For example, vertices 1 and 3 of Figure \ref{fig:small}(B) are doubly connected because the shortest path connecting them contains links in more than one layer (1--5--6--3). Vertices 5 and 6, on the other hand, are not doubly connected because the shortest path connecting them corresponds to the same layer.

A weighted multilayer network is a set of $M$ weighted networks, $\vec G=(G_1,G_2,\dots,G_M,{\mathbb E})$ where $G_{L}=(V_{L},E_{L},W_{L})$, with $L=1,2,\dots,M$; these $M$ layers are connected by weighted interlayer links ${\mathbb E}=\{ w_{x,y}\;\|\;x\in V_\alpha,\;y\in V_\beta,\;\alpha\not=\beta\}$.  

Then, the {\em dNDD} of a multilayer system must take into account the impact of the intra-layer links, on the shortest paths.
Thus, for two nodes in the same layer, $i, j\in V_L$, we define:
\begin{enumerate}
 \item $D_{w}(i,j)$, as the shortest distance over all paths connecting $i$ and $j$ in the multilayered system, being $\mathbb{P}_{i}$ its corresponding probability distribution;
 \item $D^{\beta}_w(i,j)$, the shortest distance over all paths connecting $i$ and $j$ such that at least one node in the path
 belongs to a different layer, $G_\beta$ with $\beta\not=L$, being $\mathbb{P}^\beta_i$ its corresponding probability distribution.
\end{enumerate}


For every $i\in V_L$, $\mathbb{P}_{i}$ captures the impact of the shortest paths reduction due to the multilayer, and $\mathbb{P}_i^{\beta}$ captures the impact of the paths that are forced to have, at least, one link connected to layer $G_\beta$.

In section D of the SI, is presented the construction of these probability distributions for the network depicted in Figure~\ref{fig:small}(B).

\section{Acknowledgments}
Research partially supported by Brazilian agencies FAPEMIG, CAPES, and  CNPq.
P.M.P. acknowledges support from the ``Paul and Heidi Brown Preeminent Professorship in ISE, University of Florida'', and RSF 14-41-00039 and Humboldt Research Award (Germany).
C. M. acknowledges partial support from Spanish MINECO (FIS2015-
66503-C3-2-P) and ICREA ACADEMIA.
A. D-G acknowledges financial support from MINECO via Project No. PGC2018-094754-B-C22 (MINECO/FEDER,UE) and Generalitat de Catalunya via Grant No. 2017SGR341.
M.G.R acknowledges partial support from FUNDEP.

\bibliography{bibliografia_3}

\newpage

\section{SUPPLEMENTARY INFORMATION}

\subsection{A. Simulated values for the heat model}

The heat model consider that each network vertex, $i$, has an initial chosen temperature, $x^0_i$. The vertices interact by exchanging heat in order to reach thermal equilibrium with speed proportional to the difference between their temperatures and the weight of the existing links between them. Thus, the higher the weight means the higher rate of heat transfer from one vertex to another. Therefore, a set of $N$ differential equations governing the dynamics of the system are:
$$
\frac{dx_i}{dt}=\sum_{j\not=i,j=1}^Nw_{i,j}(x_j-x_i),\;\;x_i(0)=x^0(i)\;\;\forall\;\;i=1,\dots,N
$$
which can be rewritten as:
$$
\frac{d{\mathbf x}}{dt}=-{\cal{L}}\cdot{\mathbf x},\qquad {\mathbf x}(0)={\mathbf x}^0,
$$
being ${\cal{L}}$ the graph's Laplacian matrix.

For undirected networks, this system of equations possesses a single solution given by:
$$
{\mathbf x}(t)={\mathbf P}\cdot{\mathbf E(t)}\cdot{\mathbf C},
$$
being, ${\mathbf P}$ the matrix whose columns are eigenvectors of $-{\cal{L}}$, ${\mathbf E(t)}$ the diagonal matrix in which the i-th element depends on the i-th eigenvalue of $-{\cal{L}}$, $\lambda_i$, given by $e^{\lambda_it}$ and $C$ a matrix that depends on the initial conditions given by:
$$
C={\mathbf P}^{-1}\cdot{\mathbf x}^0
$$

\subsection{B. Correlation between diffusion times computed through $\lambda$ and diffusion capacity in multiplex system.}
\begin{figure}[h]
 \centering
 \includegraphics[scale=0.4]{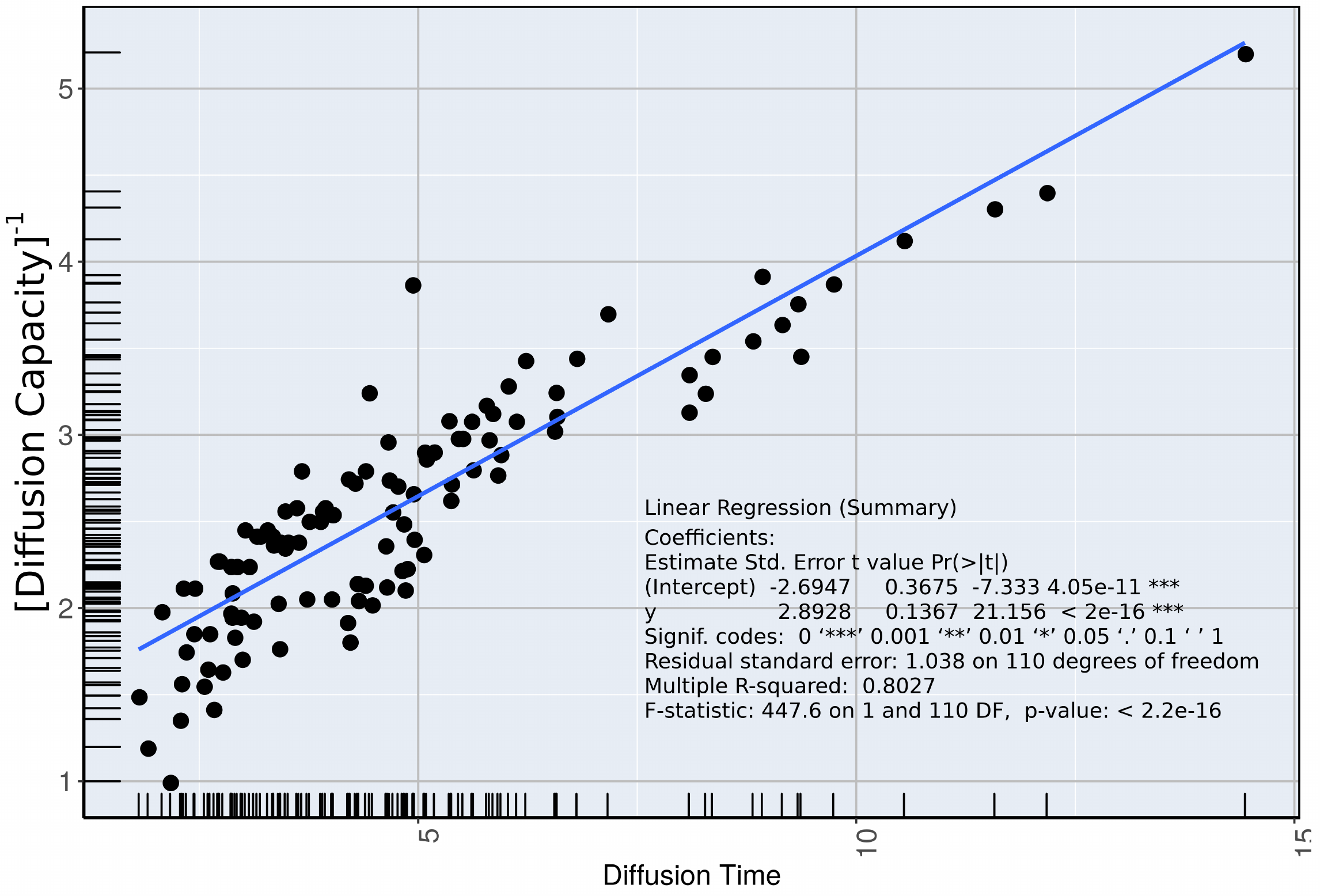}
 \caption{Diffusion capacity versus diffusion time computed through $\lambda$ of a multiplex system described in~\cite{Gomez2013}.}
 \label{fig:}
\end{figure}

\subsection{C. Numerical example of the dynamic node distance distribution in single networks.}

The computation of the dynamical node distance distribution of node 1, of the network in Figure~\ref{fig:small}, is the following:

$\mathbf{p}_{1}^+(1)=\frac{1}{3}\max \left(1-{\Delta_{1,2},0}\right)+\frac{1}{3}\max \left(1-{\Delta_{1,3},0}\right)=\frac{1}{3}\max (1-1.5,0)+\frac{1}{3}\max(1-1,0)=0$\\

$\mathbf{p}_{1}^{0}(1)=\frac{1}{3}\min \left(\Delta_{1,2},\frac{1}{\Delta_{1,2}}\right)+\frac{1}{3}\min \left(\Delta_{1,3},\frac{1}{\Delta_{1,3}}\right)=\frac{1}{3}\left(\frac{1}{1.5} +1\right)=\frac{5}{9}$\\

$\mathbf{p}_{1}^-(1)=\frac{1}{3}\max \left(1-\frac{1}{\Delta_{1,2}},0\right)+\frac{1}{3}\max \left(1-\frac{1}{\Delta_{1,3}},0\right)=\frac{1}{3}\left(\frac{1}{3} +0\right)=\frac{1}{9}$\\

$\mathbf{p}_{1}^+(2)=\frac{1}{3}\max \left(1-{\Delta_{1,4},0}\right)=\frac{1}{3}\max \left(1-0.75,0)\right)=\frac{1}{3}\left (\frac{1}{4}\right)=\frac{1}{12}$\\

$\mathbf{p}_{1}^{0}(2)=\frac{1}{3}\min \left(\Delta_{1,4},\frac{1}{\Delta_{1,4}}\right)=\frac{1}{3}\min \left(0.75,\frac{1}{0.75}\right)=\frac{1}{3}\left(0.75\right)=\frac{1}{4}$\\

$\mathbf{p}_{1}^-(2)=\frac{1}{3}\max \left(1-\frac{1}{\Delta_{1,4}},0\right)=\frac{1}{3}\max \left(1-\frac{1}{0.75},0\right)=0$\\

$\mathbf{p}_{1}^+(3)=\mathbf{p}_{1}^{0}(3)=\mathbf{p}_{1}^-(3)=\mathbf{p}_1(\infty)=0$\\

Then, \\
\begin{equation}
 \mathbb{P}_{1}=[0,\frac{5}{9},\frac{1}{9},\frac{1}{12},\frac{1}{4},0,0,0,0,0].
 \label{eq:ndw}
 \end{equation}

Considering all nodes and distances, we have the following dynamical node distance distributions:
$$
\begin{array}{l|c|c|c|c|c|c|c|c|c|c}
\mbox{Node}&\mathbf{p}^+(1)&\mathbf{p}^0(1)&\mathbf{p}^-(1)&\mathbf{p}^+(2)&\mathbf{p}^0(2)&\mathbf{p}^-(2)&\mathbf{p}^+(3)&\mathbf{p}^0(3)&\mathbf{p}^-(3)&\mathbf{p}(\infty)\\\hline
1&0&5/9&1/9&1/12&1/4&0&0&0&0&0\\
2&1/6&7/18&1/9&1/6&1/6&0&0&0&0&0\\
3&1/3&2/3&0&0&0&0&0&0&0&0\\
4&1/6&1/6&0&1/4&5/12&0&0&0&0&0\\\hline
\end{array}
$$

\subsection{D. Numerical example of the dynamical node distance distribution for multilayer networks.}

In layer $\alpha$, node one posses a direct link to nodes 2 and 3, and two link to node 4, then, the corresponding geodesic distances are $D_g(1,2)=D_g(1,3)=1$, and  $D_g(1,4)=2$.
Now, we compute the minimum weighted distances $D_w$ to each node from node 1. For example, for $D_w(1,2)$, we have three possibilities, $D_w(1\rightarrow 2)=1/0.5=2$, $D_w(1 \rightarrow 3 \rightarrow 2)=1+1/2=3/2$, and $D_w(1 \rightarrow 5 \rightarrow 6 \rightarrow 4 \rightarrow 3 \rightarrow 2)= 1/10+1/10+1/10+1/2+1/2=13/10$. Then, $D_w(1,2)=13/10$.
In the same way we compute $D_w(1,3)=8/10$, and  $D_w(1,4)=3/10$, and the corresponding $\Delta_{1,2}=13/10$, $\Delta_{1,3}=8/10$, and $\Delta_{1,4}=3/20$.

$$
\mathbf{p}_{1}^+(1)=\frac{1}{3}\max \left(1-{\Delta_{1,2},0}\right)+\frac{1}{3}\max \left(1-{\Delta_{1,3},0}\right)=\frac{1}{3}. \left(0+\frac{2}{10}\right)=\frac{1}{15}=0.067
$$

$$
\mathbf{p}_{1}^{0}(1)=\frac{1}{3}\min \left(\Delta_{1,2},\frac{1}{\Delta_{1,2}}\right)+\frac{1}{3}\min \left(\Delta_{1,3},\frac{1}{\Delta_{1,3}}\right)=\frac{1}{3}.  \left(\frac{10}{13}+\frac{8}{10}\right)= 0.523
$$

$$
\mathbf{p}_{1}^-(1)=\frac{1}{3}\max \left(1- \frac{1}{\Delta_{1,2}},0\right)+\frac{1}{3}\max \left(1- \frac{1}{\Delta_{1,3}},0\right)=0.077
$$

$$
\mathbf{p}_{1}^+(2)=\frac{1}{3}\max \left(1-{\Delta_{1,4},0}\right)= \frac{1}{3} 0.85=0.283
$$

$$
\mathbf{p}_{1}^{0}(2)=\frac{1}{3}\min \left(\Delta_{1,4},\frac{1}{\Delta_{1,4}}\right)=\frac{1}{20}=0.05
$$

$$
\mathbf{p}_{1}^-(2)=\frac{1}{3}\max \left(1- \frac{1}{\Delta_{1,4}},0\right)=0
$$
$$
\mathbf{p}_{1}^+(3)=\mathbf{p}_{1}^{0}(3)=\mathbf{p}_{1}^-(3)=\mathbf{p}_1(\infty)=0
$$

Then,
\begin{equation}
\label{1-multi}
\mathbb{P}_{\mbox{1,m}}=[0.067,0.523,0.077,0.284,0.05,0,0,0,0,0]
\end{equation}
Repeating the process for all nodes and distances, we have the {\em multilayer distance distributions}.

Now, considering paths that forcibly passes through layer $\beta$, $D_w(1,2)=\min\{D_w(1,5)+D_w(5,2),D_w(1,6)+D_w(6,2)\}=0.8$, $D_w(1,3)=\min\{D_w(1,5)+D_w(5,3),D_w(1,6)+D_w(6,3)\}=0.3$ and $D_w(1,4)=\min\{D_w(1,5)+D_w(5,4),D_w(1,6)+D_w(6,4)\}=0.8$. Then, $\Delta_{1,2}=0.8$, $\Delta_{1,3}=0.3$, and $\Delta_{1,4}=0.4$.

$$
\mathbf{p}_{1}^+(1)=\frac{1}{3}\max \left(1-{\Delta_{1,2},0}\right)+\frac{1}{3}\max \left(1-{\Delta_{1,3},0}\right)=\frac{1}{3}. \left(\frac{2}{10}+\frac{7}{10}\right)=\frac{3}{10}=0.3
$$

$$
\mathbf{p}_{1}^{0}(1)=\frac{1}{3}\min \left(\Delta_{1,2},\frac{1}{\Delta_{1,2}}\right)+\frac{1}{3}\min \left(\Delta_{1,3},\frac{1}{\Delta_{1,3}}\right)=\frac{1}{3}.  \left(\frac{8}{10}+\frac{3}{10}\right)=0.367
$$
$$
\mathbf{p}_{1}^-(1)=\frac{1}{3}\max \left(1- \frac{1}{\Delta_{1,2}},0\right)+\frac{1}{3}\max \left(1- \frac{1}{\Delta_{1,3}},0\right)=0
$$
$$
\mathbf{p}_{1}^+(2)=\frac{1}{3}\max \left(1-{\Delta_{1,4},0}\right)= \frac{1}{3} \frac{6}{10}= 0.2
$$

$$
\mathbf{p}_{1}^{0}(2)=\frac{1}{3}\min \left(\Delta_{1,4},\frac{1}{\Delta_{1,4}}\right)= \frac{1}{3} \frac{4}{10}= 0.133
$$

$$
\mathbf{p}_{1}^-(2)=\frac{1}{3}\max \left(1- \frac{1}{\Delta_{1,4}},0\right)=0
$$
$$
\mathbf{p}_{1}^+(3)=\mathbf{p}_{1}^{0}(3)=\mathbf{p}_{1}^-(3)=\mathbf{p}_1(\infty)=0
$$

\begin{equation}
\label{1-multi}
\mathbb{P}^{\beta}_{\mbox{1}}=[0.3,0.367,0,0.2,0.133,0,0,0,0,0]
\end{equation}
Repeating the process for all nodes and distances, we have the distance distribution in which shortest paths are forced to use all layers.

\subsection{E. The cumulative Jensen-Shannon difference (CDD)}

Here we consider the Node Distance Distribution in unweighted networks. Thus, for each vertex $x$, $P_x=(p_1,p_2,\dots,p_{N-1},p\infty)$ gives the vector of the density of nodes at a geodesic distance from $x$. For example, $p_1$ gives the fraction between the number os nodes at a distace 1 from $x$ and $(N-1)$, being $N$ the number of vertices in the network.

The Jensen-Shannon divergence was successfully used in computing the distance between the intricate patterns of connectivity given by the Node Distance Distributions. However, diffusive systems must carefully consider not only the connectivity patterns but also the appropriate way in which it connects throughout the network.

\begin{figure}[h]
 \centering
 \includegraphics[scale=0.7]{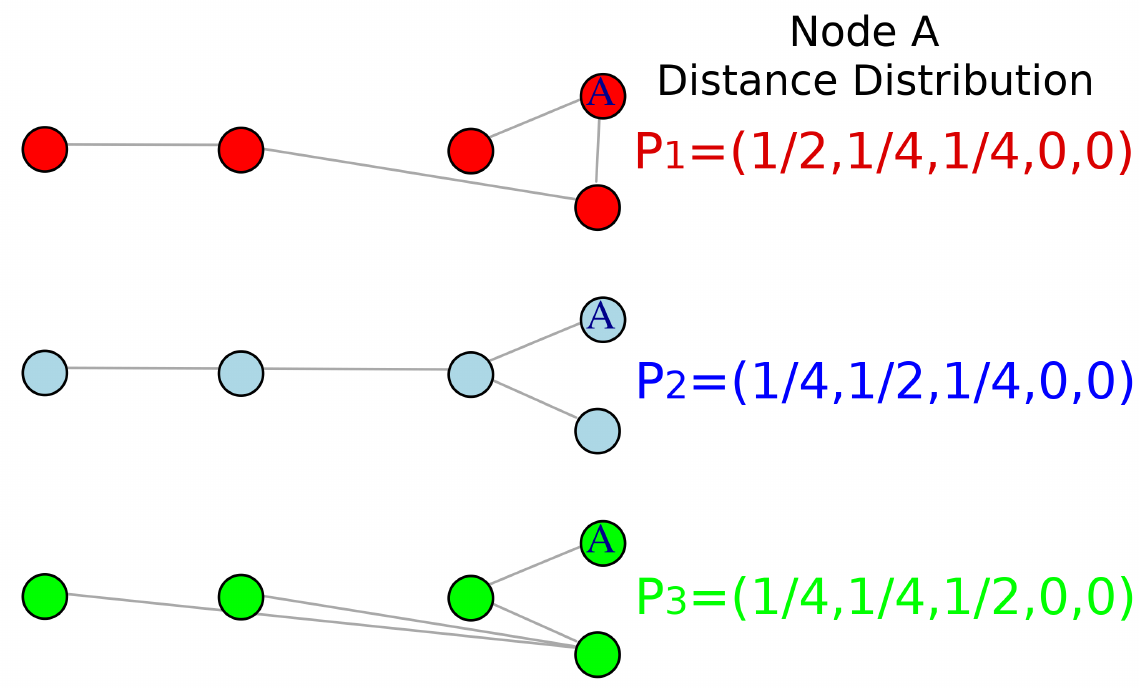}
 \caption{Node distance distribution for node A in three different unweighted networks.}
 \label{fig:cumulativa_small}
\end{figure}

Consider networks in Figure~\ref{fig:cumulativa_small}. By comparing the difference in connectivity patterns from node A via the Jensen-Shannon divergence applied to the NDD, we obtain $JS(P_1,P_2)=JS(P_1,P_3)=JS(P_2,P_3)$. This result shows the inefficiency of the divergence in adequately measuring the fundamental difference in connectivity patterns of node A. Note that for a random walker starting from the vertex A, with 1 step it can reach 50\% of all possible nodes considering $P_1$ and 25\% for both $P_2$ and $P_3$. However, with two steps, both $P_1$ and $P_2$ reach 75\% of the possible nodes, while the third reaches only 50\%. With three steps, a random walker can achieve 100\% of the vertices. In this way, it makes sense considering the distance from the first to the second to be smaller than the first to the third.

To overcome this problem we propose the use of a cumulative JS measure like. Let $P=(p_1,p_2,\dots,p_N)$ a discrete probability distribution, and let $P^c(x)$ defined as
$$
P^c(x)=(\sum_{i=1}^xp_i,p_{x+1},p_{x+2},\dots,P_N).
$$
We can check that $P^c(1)=P$. For the probability distributions $P_1$, $P_2$ and $P_3$ in Figure~\ref{fig:cumulativa_small}, for example we have:
$$
\begin{array}{ccc}
P_1^c(2)=(1/2+1/4,1/4,0,0)&P_1^c(3)=(1/2+1/4+1/4,0,0)\\
P_2^c(2)=(1/4+1/2,1/4,0,0)&P_2^c(3)=(1/4+1/2+1/4,0,0)\\
P_3^c(2)=(1/4+1/4,1/2,0,0)&P_1^c(3)=(1/4+1/4+1/2,0,0)
\end{array}
$$
The interpretation of this cumulative-like distribution is that the first number of the vector $P^c(x)$ represents the fraction of nodes at a distance at most $x$, the second the fraction of nodes at a distance $x+1$ and so on.

It is easy to see that
$$
JS(P^c(x),Q^c(x))\leq JS(P^c(y),Q^c(y))\;\;\forall\;\;x\geq y
$$

We define the Cumulative Distance difference between two NDD by the measure:
$$
CDD(P,Q)=\sum_{x}JS(P^c(x),Q^c(x))
$$

CDD is a metric and here JS is the square root of the normalized JS divergence.

For the network in Figure~\ref{fig:cumulativa_small} we obtain that $CDD(P_1,P_2)<CDD(P_1,P_3)$.

\section{E. Heatmaps for the experiment in Figure 5}

\begin{figure}
  \centering
  \includegraphics[width=\linewidth]{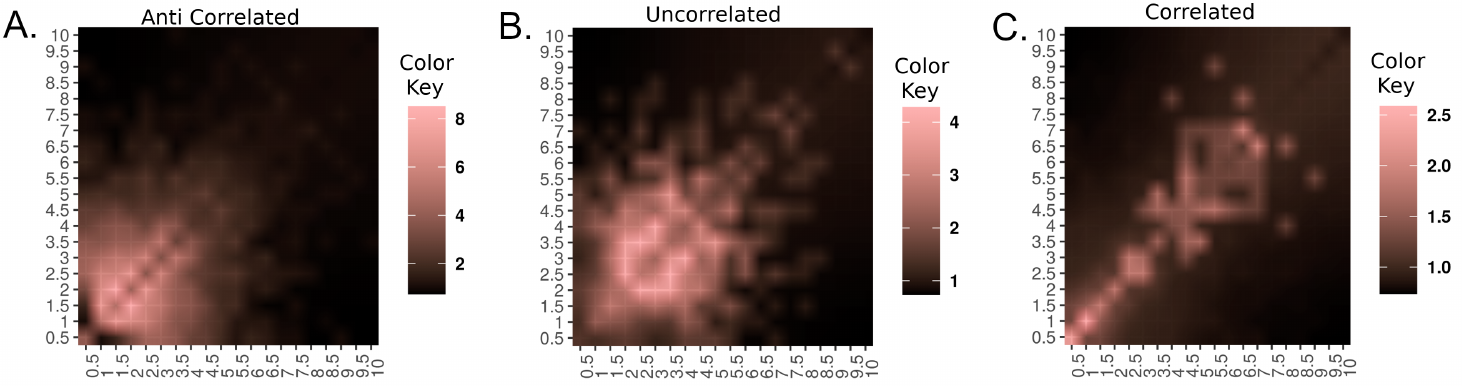}
  \caption{Heatmaps for the experiment in Figure 5.}
\end{figure}

\subsection{F. Values}

\begin{table}[h]
{%
\newcommand{\mc}[3]{\multicolumn{#1}{#2}{#3}}
\begin{center}
\begin{tabular}{l|l|lllllllll}\cline{2-10}
 & \mc{9}{c|}{Node} &  \\\hline
\mc{1}{|l|}{Diffusion Capacity} & \mc{1}{l|}{1} & \mc{1}{l|}{2} & \mc{1}{l|}{3} & \mc{1}{l|}{4} & \mc{1}{l|}{5} & \mc{1}{l|}{6} & \mc{1}{l|}{7} & \mc{1}{l|}{8} & \mc{1}{l|}{9}& \mc{1}{l|}{Average}\\\hline
\mc{1}{|l|}{$\Lambda_x(G_1)$} & \mc{1}{l|}{0.2825} & \mc{1}{l|}{0.2825} & \mc{1}{l|}{0.2825} & \mc{1}{l|}{0.2825} & \mc{1}{l|}{0.2825} & \mc{1}{l|}{0.2825} & \mc{1}{l|}{0.2825} & \mc{1}{l|}{0.2825} & \mc{1}{l|}{1.0000} & \mc{1}{l|}{0.3622} \\\hline
\mc{1}{|l|}{${\cal M}_x(G_1)$ in S1} & \mc{1}{l|}{\color{red}0.2909} & \mc{1}{l|}{\color{red}0.2867} & \mc{1}{l|}{\color{red}0.2867} & \mc{1}{l|}{\color{red}0.2867} & \mc{1}{l|}{\color{red}0.2867} & \mc{1}{l|}{\color{red}0.2867} & \mc{1}{l|}{\color{red}0.2867} & \mc{1}{l|}{\color{red}0.3074} & \mc{1}{l|}{\color{blue}0.9950} & \mc{1}{l|}{0.3682} \\\hline
\mc{1}{|l|}{${\cal M}_x(G_1)$ in S2} & \mc{1}{l|}{\color{red}0.2912} & \mc{1}{l|}{\color{red}0.2912} & \mc{1}{l|}{\color{red}0.2912} & \mc{1}{l|}{\color{red}0.2912} & \mc{1}{l|}{\color{red}0.2912} & \mc{1}{l|}{\color{red}0.2912} & \mc{1}{l|}{\color{red}0.2912} & \mc{1}{l|}{\color{red}0.2912} & \mc{1}{l|}{\color{blue}0.9999} & \mc{1}{l|}{0.3699}\\\hline
\mc{1}{|l|}{$\Lambda_x(G_2)$} & \mc{1}{l|}{0.3400} & \mc{1}{l|}{0.3103} & \mc{1}{l|}{0.3103} & \mc{1}{l|}{0.3103} & \mc{1}{l|}{0.3103} & \mc{1}{l|}{0.2321} & \mc{1}{l|}{0.2506} & \mc{1}{l|}{0.5657} & \mc{1}{l|}{0.5657} & \mc{1}{l|}{0.3550}\\\hline
\mc{1}{|l|}{${\cal M}_x(G_2)$ in S1} & \mc{1}{l|}{\color{blue}0.3396} & \mc{1}{l|}{\color{blue}0.3100} & \mc{1}{l|}{\color{blue}0.3100} & \mc{1}{l|}{\color{blue}0.3100} & \mc{1}{l|}{\color{blue}0.3100} & \mc{1}{l|}{\color{red}0.2379} & \mc{1}{l|}{\color{red}0.2537} & \mc{1}{l|}{\color{blue}0.5652} & \mc{1}{l|}{\color{red}0.5914} & \mc{1}{l|}{0.3587}\\\hline
\mc{1}{|l|}{$\Lambda_x(G_3)$} & \mc{1}{l|}{0.2466} & \mc{1}{l|}{0.2466} & \mc{1}{l|}{0.2696} & \mc{1}{l|}{0.2696} & \mc{1}{l|}{0.2696} & \mc{1}{l|}{0.1854} & \mc{1}{l|}{0.1854} & \mc{1}{l|}{0.2162} & \mc{1}{l|}{0.4160} & \mc{1}{l|}{0.2561}\\\hline
\mc{1}{|l|}{${\cal M}_x(G_3)$ in S2} & \mc{1}{l|}{\color{red}0.2511} & \mc{1}{l|}{\color{red}0.2511} & \mc{1}{l|}{\color{red}0.2732} & \mc{1}{l|}{\color{red}0.2732} & \mc{1}{l|}{\color{red}0.2732} & \mc{1}{l|}{\color{red}0.1921} & \mc{1}{l|}{\color{red}0.1921} & \mc{1}{l|}{\color{red}0.2231} & \mc{1}{l|}{\color{red}0.4438} & \mc{1}{l|}{0.2636}\\\hline
\end{tabular}
\end{center}
}%
\caption{Mono and multilayer diffusion capacity values for the systems depicted in Figure \ref{fig:smallest}.}
 \label{fig:tabela}
\end{table}

\subsection{F. Diffusion Times and Diffusion Capacity}

Diffusion time $\tau$ can be defined as proportional to its diffusion capacity, adjusted by a real decreasing function $f$, as $\tau=f({\cal M}(\vec G))$. This approach allows the adjustment of specific characteristics of the dynamical process, such as, infection or recovery times, etc.

\begin{figure}
  \centering
  \includegraphics[width=5cm]{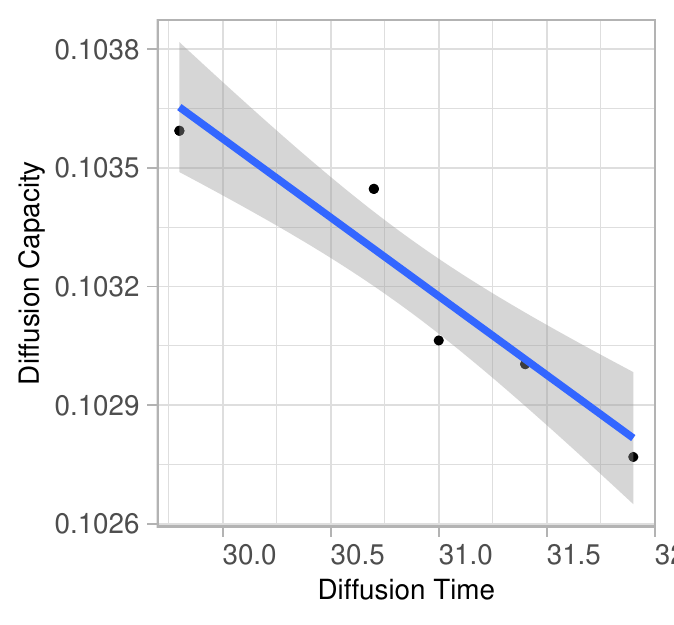}
  \caption{The linear fit between diffusion capacity and diffusion time for different Heat Model initial conditions on the network in Figure 1.}\label{decreasing}
\end{figure}

As an example, consider the heat model with different initial conditions with the perturbation of only one vertex initially. The diffusion time is taken as the time needed to the standard deviation of the temperature to be less than $10^{-6}$. Figure \ref{decreasing} shows a linear fit between diffusion time and diffusion capacity at the beginning of the process.

\subsection{G. Diffusion Times and Diffusion Capacity}
\begin{figure}
  \centering
  \includegraphics[width=\linewidth]{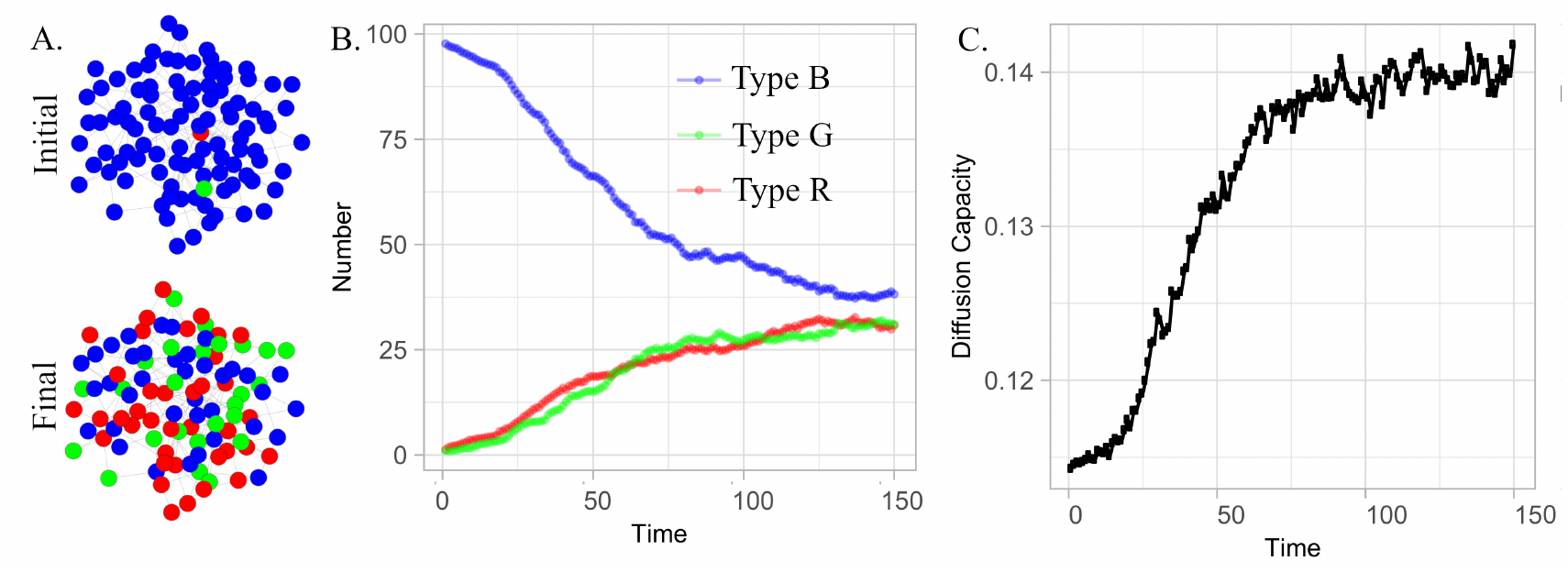}
  \caption{Competing dynamics with heterogeneous types of nodes. In A it is depicted a network in its initial and final state in which a node can change the color of its neighbor with a probability $p$. The number of nodes of each color in time, and the network diffusion capacity are shown in B and C, respectively.}
  \label{id1}
\end{figure}

The following experiment is an example of competing dynamics in single and multilayer networks which shows how heterogeneity increases the diffusion capacity. Consider a network with three types of nodes: red (R), green (G), and blue (B). Vertices that are directly connected can change the color of their neighbors with a fixed probability $p$. Figure \ref{id1}-A shows a small network in which $p$ is the weight of links. Initially, the blue nodes are the majority, and only node red and one green node are present, then, most nodes are connected to blue nodes. As the process evolves, more nodes are being connected to nodes of different colors, and connectivity becomes more heterogeneous Figure \ref{id1}-B. This grow in heterogeneity increases the speed at which the system structure changes, as more pairs of nodes change colors. Figure \ref{id1}-C shows how the average diffusion capacity grows as heterogeneity increases.

\begin{figure}
  \centering
  \includegraphics[width=.8\linewidth]{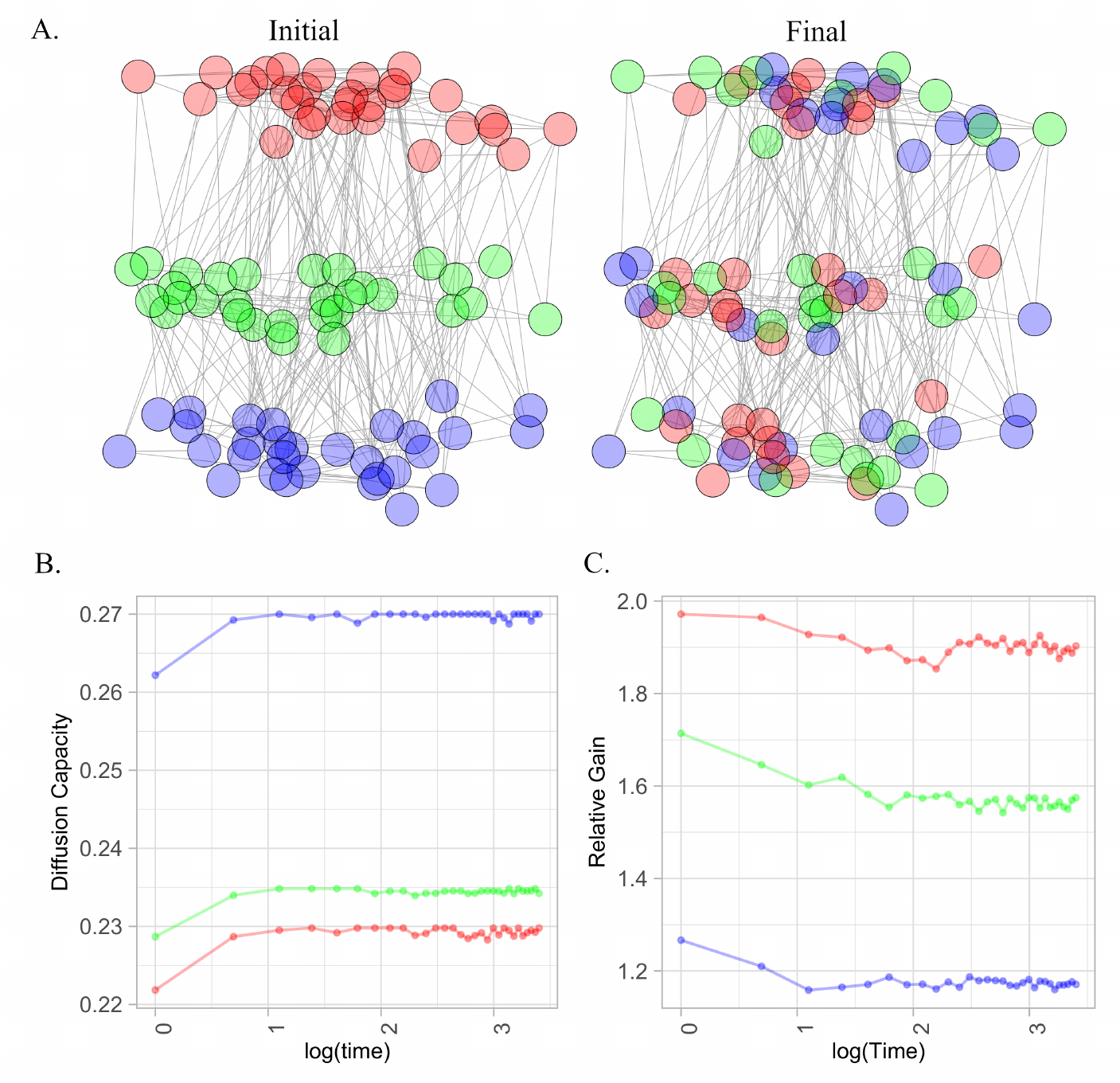}
  \caption{Multilayer network composed of three layers, in which each one is composed exclusively of one color. In A it is depicted a multilayer network in its initial and final state in which a node can change the color of its neighbor with a probability $p$. Values of diffusion capacity of each isolated layers, and their corresponding relative gains are shown in B and C, respectively.}
  \label{id2}
\end{figure}

A similar behavior is found when this model is applied to a multilayer system in which each layer is composed by one color nodes, as shown in Figure \ref{id2}-A. In this case, the diffusion capacity of each layer increases with time as the interaction between layers grows, and then stabilizes  (Figure \ref{id2}-B). The isolated layer with lower diffusion capacity is the one that gains the most with the interaction with the other layers. As the interaction between layers grows, the relative gain of each layer decreases as the dynamics within each layer becomes more efficient, and the interlayer paths loose importance (Figure \ref{id2}-C).


\end{document}